\documentclass[twocolumn,prl,superscriptaddress,aps,amsmath,floatfix,longbibliography]{revtex4-2}

\usepackage[pdfborder={0 0 0},colorlinks=true,linkcolor=blue,urlcolor=blue,citecolor=blue]{hyperref}

\usepackage{graphics}
\usepackage{graphicx}
\usepackage{epstopdf}
\usepackage{dcolumn}
\usepackage{bm}
\usepackage{longtable}
\usepackage{epsfig}
\usepackage{times}
\usepackage{url}

\usepackage{color}

\usepackage{color,soul}

\begin{document}
\title{Gravitational photon echo}

\author{Wen-Te \surname{Liao}}
\email{wente.liao@g.ncu.edu.tw}
\affiliation{Department of Physics, National Central University, Taoyuan City 32001, Taiwan}
\affiliation{Physics Division, National Center for Theoretical Sciences, Taipei 10617, Taiwan}
\affiliation{Quantum Technology Center, National Central University, Taoyuan City 32001, Taiwan}

\author{Sven \surname{Ahrens}}
\email{ahrens@shnu.edu.cn}
\affiliation{Shanghai Normal University, Shanghai 200234, China} 
\date{\today}
\begin{abstract}
The generation, controls, and storage of the gravitationally induced photon echo using the 8.4 eV Thorium-229 nuclear clock transition on Earth are theoretically investigated.
With its exceptionally narrow linewidth  of approximately 1 mHz and  high quality factor in the order of  $10^{19}$, the Thorium-229 clock transition allows for the potential detection of gravitational-redshift effects at millimeter-scale altitude variations.
Moreover, the 1740-second half lifetime of the Thorium-229 isomeric state allows for the slow movement of the system within its coherence time.
Along this line, we explore the generation, controls, and storage of  photon echo arising from  a rotation induced inversion of the gravitational frequency shift in a single target or  in multiple equally spaced samples.
 Our approach lays the foundation for the controllable gravitational quantum optics on Earth.
\end{abstract}

\keywords{quantum optics,interference effect}
\maketitle
%
%
Einstein’s general relativity predicts gravitational redshift, which has been observed on Earth over distances from millimeters to global scales \cite{Pound1959, Potzel1992, Chou2010, Bothwell2022}. 
However, advancing this effect beyond passive observations toward an active control or practical applications in coherent quantum systems remains a challenge \cite{Zych2011, Liao2015, Lee2025}. This requires a quantum system that meets two key criteria:
(1) sensitivity to gravitational redshift on Earth, and
(2) coherence time exceeding the control duration.
In this Letter, we propose a solution using the Thorium-229 ($^{229}$Th) nuclear clock transition to generate, control, and store gravitationally induced photon echo (GPE). 
With its 8.4 eV excitation energy within the vacuum ultraviolet (VUV) region, 1740 s isomer half-life, and a quality factor about $10^{19}$, $^{229}$Th   is expected to demonstrate the millimeter-scale sensitivity to gravitational redshift and meets the above first criterion \cite{Peik2003, Seiferle2019, Sikorsky2020, Beeks2021, Peik2021, Kraemer2023, Tiedau2024, Liao2015, Lee2025}.
Photon echoes arise from controllable rephasing mechanisms or the periodic spectral beating of coherent ensembles \cite{Afzelius2009, De2008, Afzelius2010, Main2021, Helisto1991, Vagizov2014}.
In $^{229}$Th-doped targets, the gravitational redshift induces inhomogeneous spectral broadening or  frequency shifts between vertically separated targets. Moreover, the long $^{229m}$Th lifetime meets the above second criterion and makes  rotations of the VUV-excited system within its coherence time possible, dynamically altering the gravitational gradient and leading to the rephasing.
These spectral effects, arising from Earth's gravity and the unique properties of the $^{229}$Th clock transition, enable the realization of GPEs. In this work, we investigate two types of GPE:
(i) A gravitational gradient photon echo (GGPE) from a single $^{229}$Th-doped target, where reversible gravitational broadening rephases nuclear polarization, reminding gradient echo memory techniques \cite{Hetet2008, Hosseini2009, Hetet2008b, Liao2014b};
(ii) A gravitational frequency comb (GFC) formed by a vertical array of equally spaced $^{229}$Th-doped targets, where altitude-dependent gravitational redshift naturally constitutes a frequency comb structure without the need for conventional spectral hole burning \cite{Afzelius2009, De2008, Afzelius2010, Main2021, Zhang2019, Velten2024}.
Our scheme potentially offers applications in relativistic geodesy \cite{Flury2016}, e.g., precision measurements of gravitational potential at the millimeter scale.

%
%
%

\begin{figure}[b]
\includegraphics[width=0.48\textwidth]{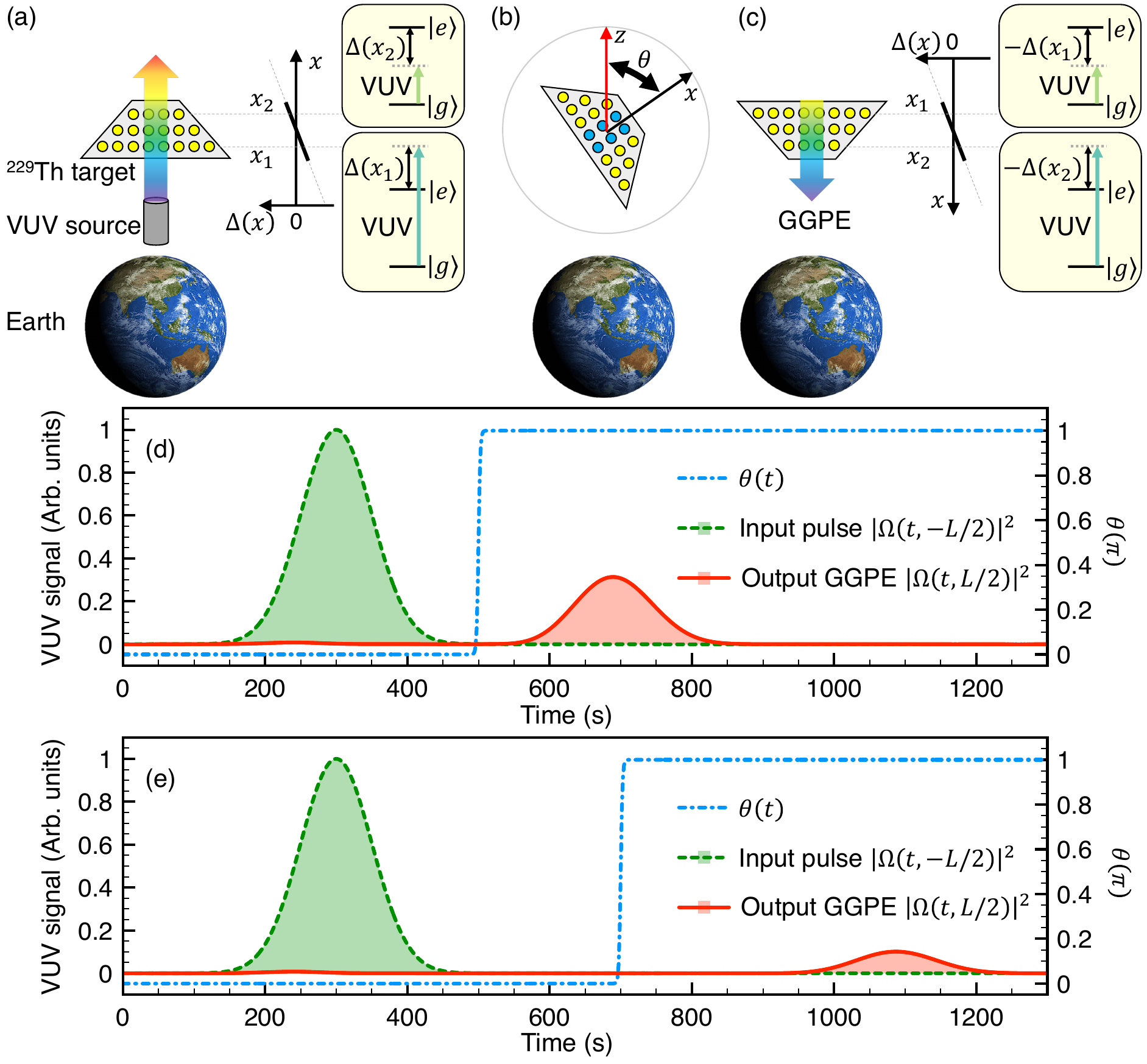}
\caption{\label{fig1} 
(a) A VUV pulse (rainbow-colored upward arrow) illuminates a $^{229}$Th target (gray trapezoid with yellow dots) from below. The detuning $\Delta\left( x\right) $ indicates the vertical variation in gravitational redshift. The upper (lower) yellow panel shows excitation of the two-level nuclear transition $\vert g\rangle \rightarrow \vert e\rangle$ at the top $x = x_2$ (bottom $x = x_1$) with detuning $\Delta\left[ x_2 \right] $ ($\Delta\left[ x_1 \right] $). 
The shape of each target serves for indicating its orientation.
(b) After illumination, the $^{229}$Th target is rotated by a polar angle $\theta$. Yellow (blue) dots represent nuclei in the ground (isomeric) state.
(c) When the target is rotated upside down ($\theta = \pi$), it emits a GGPE.
(d) and (e) VUV signals following the inversion of the target at $t = 500$ s and $t = 700$ s, respectively. The green-dashed-filled line shows the input VUV pulse $\vert \Omega(t, -L/2) \vert^2$, the red-solid-filled line the output GGPE  $\vert \Omega(t, L/2) \vert^2$, and the blue dash-dotted line the rotation angle $\theta(t)$.
}
\end{figure}
The coupling between VUV light and two-level $^{229}$Th nuclei, comprising ground state $\vert g\rangle$ ($^{229g}$Th) and isomeric state $\vert e\rangle$ ($^{229m}$Th) in Fig.~\ref{fig1}(a), is governed by the optical Bloch equations in the perturbative regime \cite{Grynberg2010, Liao2012b, Lin2022, Chen2022}:
\begin{equation}
\partial_{t}\rho_{eg}^{\left( n\right) }  = -\left[\frac{\Gamma_0}{2}+\gamma+i\Delta_n\left( z_n, x_n, \theta \right) \right] \rho_{eg}^{\left( n\right) }+\frac{i}{2}\Omega_n ,\label{eq1}
\end{equation}
\begin{equation}
\frac{1}{c}\partial_{t} \Omega_n + \partial_{x_n} \Omega_n = i\eta\rho_{eg}^{\left( n\right) }  . \label{eq2}
\end{equation}
Here index $n$ indicates the quantity for the $n$th target in an array, and we will remove the index $n$  in a single-target system.
$\rho_{eg}^{\left( n\right) } $  is the nuclear polarization, i.e., the off-diagonal element of the  two-level  density matrix. 
$\Omega_n$ is the VUV Rabi frequency and proportional to the  electric field strength of the VUV light \cite{Palffy2008}.
$\Gamma_0 = \ln2/1740$ rad$\cdot$Hz is the natural  linewidth of the nuclear clock transition $\vert g\rangle\rightarrow\vert e\rangle$ \cite{Tiedau2024}, and $\gamma = \ln2/630$ rad$\cdot$Hz is the enhanced decoherence rate in a $^{229}$Th:CaF$_2$ crystal \cite{Tiedau2024}. $\eta = \Gamma_0 \xi/\left( 2 L\right) $ is the light-nucleus coupling constant, where $\xi$ and $L$ are respectively the optical depth and the thickness of a target.
The gravitational detuning \cite{Liao2015, Lee2025}
\begin{equation}\label{eq3}
\Delta_n\left( z_n, x_n, \theta \right) \simeq  - \frac{E_t G M_E}{\hbar c^2 R_E^2}\left( z_n+x_n\right) \cos\theta
\end{equation}
is the difference between the nuclear transition frequency, gravitational redshift, and the frequency of a VUV light  \cite{Pound1959, Chou2010,  Liao2015,  Bothwell2022}.
Here
$R_E$ is the average radius of the Earth,
$M_E$ is the mass of the Earth,
$G$ is the gravitational constant,
$E_t = 8.4$ eV is the nuclear clock transition energy \cite{Tiedau2024},  
and
$c$ is the speed of light in vacuum.
For the description of the  $n$th target, we introduce two coordinates: 
$-L/2\leq x_n \leq L/2$, which is aligned along the light propagation direction, and 
$z_n$, which represents the initial altitude of the center of the $n$th target relative to the system's center.
For convenience, the central frequency of the VUV pulse is always tuned to resonate with the clock transition at the center of the system, where the gravitational redshift is compensated.
Later we will rotate the whole system around  its center by a polar angle $\theta$, and so the gravitational detuning at the $n$th target is proportional to the projection $\left( z_n+x_n\right) \cos\theta$.
In the time sequence, we shine a VUV Gaussian pulse with a duration of $\tau_s$ on  the first target  at $t=t_0$  and treat the output field $\Omega_{n-1}\left( t, L/2 \right) $ from the $\left( n-1\right) $th target as the input $\Omega_{n}\left( t, -L/2 \right) $ of the $n$th target.
The very short light propagation time on the order of nanoseconds between the adjacent samples is neglected in our model \cite{Zhang2019}.
Accordingly, the boundary conditions are $\Omega_1 \left( t, -L/2\right) = \exp \left[ -\left( \frac{t-t_0}{\tau_s}\right) ^2 \right] $ and $\Omega_{n\geq 2} \left( t, -L/2\right) = \Omega_{n-1} \left( t, L/2\right)$. 
The initial conditions are $\rho_{eg}^{\left( n\right) }\left( 0, x_n\right) = 0$ and  $\Omega_n \left( 0, x_n\right) = 0$, i.e., neither nuclear polarization  nor VUV field pre-exists in the  system.

Figure~\ref{fig1}(a) shows a single $^{229}$Th-doped target illuminated by an upward-propagating VUV pulse, which experiences a gravitational redshift along its path. The black curve $\Delta\left( x\right)$ represents the initial detuning gradient due to gravity. This spatial variation enables a gradient photon echo, where each spectral component of the pulse is absorbed at its resonant position. To retrieve the echo, the gradient must be reversed \cite{Hetet2008, Hosseini2009, Hetet2008b, Liao2014b}. Owing to the long lifetime of the $^{229m}$Th isomeric state, this can be done via a slow rotation of the target by a polar angle $\theta$, as shown in Fig.\ref{fig1}(b). Yellow (blue) dots denote nuclei in the $^{229g}$Th ground ($^{229m}$Th isomeric) state. Fig.~\ref{fig1}(c) illustrates the case of $\theta = \pi$, i.e., the target is turned upside down.
By numerically solving Eq.\eqref{eq1} and Eq.\eqref{eq2} with parameters $\left( \xi, L, \tau_s \right) = \left( 386, 4.8 \ \mathrm{cm}, 100 \ \mathrm{s} \right)$, we obtain Fig.~\ref{fig1}(d) and (e) showing GGPE signals from target inversions at $t=500$ s and $t=700$ s, respectively. 
A VUV Gaussian pulse (green-dashed-filled line) impinges on the target at $t=300$ s, followed by a rotation $\theta\left( t\right) $ (blue-dashed-dotted line). The GGPE signal (red solid) appears at (d) $t=700$ s and (e) $t=1100$ s, mirroring the input pulse about the switching time and manifesting the time reversal mechanism \cite{Hetet2008, Hosseini2009, Hetet2008b, Liao2014b}.
Comparing the $\Delta\left( x\right) $ curve before (in Fig.\ref{fig1}(a)) and after (in Fig.\ref{fig1}(c)) the target inversion, one can observe that the detuning gradient is reversed. The gradient reversion leads  to nuclear rephasing and GGPE \cite{Hetet2008, Hosseini2009, Hetet2008b, Liao2014b}. 
Moreover, the decay of the GGPE signal is dominated by the decoherence from the target environment, with a coherence time of 630 s in the Th:CaF2 crystal \cite{Tiedau2024}. 
The GGPE signal in (e) is weaker than that in (d), because  $\rho_{eg}$ experiences the decoherence for a longer period in (e) than (d).
The vertical target thickness $L$ sets the maximum  absorption bandwidth of a single target.  
An optimal $L = 4.8$ cm provides an absorption linewidth of $116\Gamma_0$, which fully covers the  bandwidth of the input pulse. However, technical challenges may arise with growing a very thick target for having an even broader absorption bandwidth. To overcome this thickness-bandwidth problem, we next propose using multiple equally spaced thin targets to form a GFC.

\begin{figure}[b]
\includegraphics[width=0.48\textwidth]{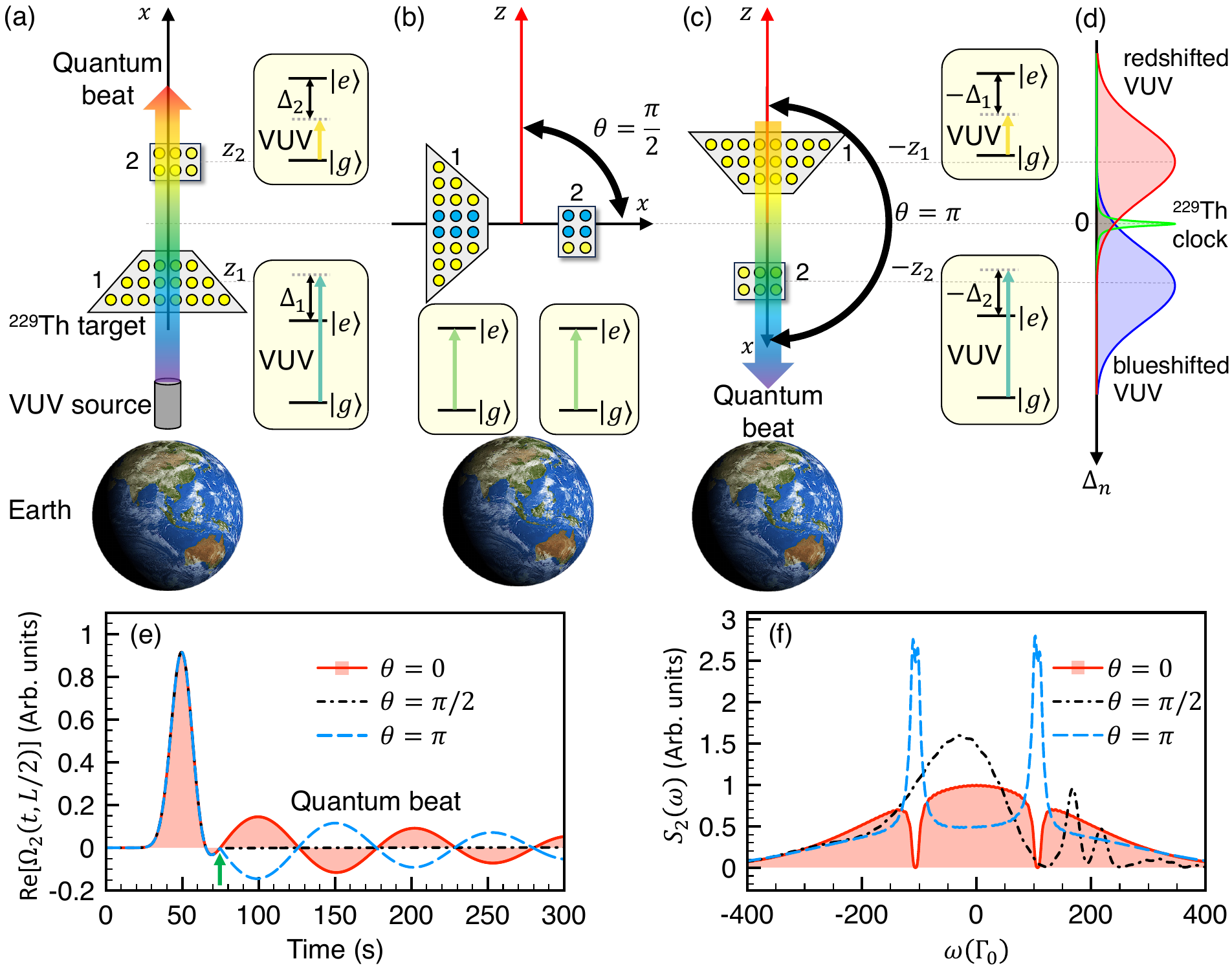}
\caption{\label{fig2}
(a) A VUV pulse (rainbow-colored upward arrow) illuminates two $^{229}$Th targets (gray trapezium at $z_1$ and square at $z_2$) separated by 8.8 cm. The VUV is resonant with the nuclear clock transition at the midpoint.
(b) When the targets are rotated by $\theta = \pi/2$ around their midpoint, they reach equal altitude, equalizing the gravitational redshift and suppressing coherent emission.
(c) The upside-down configuration of $\theta = \pi$. Target shapes indicate orientation.
(d) Spectra in the $n$th nuclear frame. 
Red (blue) line depicts the red (blue)-shifted VUV spectrum, and green line represents the $^{229}$Th clock resonance.
(e) The real part of the output VUV field Re$\left[ \Omega_2\left( t, L/2\right) \right]$, and
(f) spectrum $S_2\left( \omega\right) $.
Red-solid-filled, black-dashed-dotted, and blue-dashed lines correspond to $\theta = 0$, $\pi/2$, and $\pi$, respectively. Rotation occurs at $t = 74.8$ s indicated by the green-upward arrow in (e). 
}
\end{figure}
%
We start with a two-target system as shown in Fig.\ref{fig2}(a), target 1 (gray trapezium) at $z_1 = -4.4$ cm with detuning $\Delta_1 = 106.6\Gamma_0$, and target 2 (gray square) at $z_2 = 4.4$ cm with $\Delta_2 = -106.6\Gamma_0$, are vertically aligned. Target shapes are for identification only.
Figure~\ref{fig2}(b) shows the case of a $\theta = \pi/2$ rotation after VUV excitation, bringing both targets to resonance with zero detuning.
In the inversion case, i.e., $\theta = \pi$, illustrated by Fig.~\ref{fig2}(c), two targets exchange positions, flipping the signs of $\Delta_1$ and $\Delta_2$.
Figure\ref{fig2}(d) shows the spectral structure in the local reference frame of the $n$th target, which we term the "$n$th nuclear frame" for brevity. 
As a VUV pulse propagates through the lower (upper) target, its spectrum appears blue (red)-shifted relative to the $^{229}$Th resonance (green line), causing nuclear absorption on both shoulders of the input VUV spectrum.
We numerically solve Eq.\eqref{eq1} and Eq.\eqref{eq2} for the two-target system using parameters $\left( \xi, L, t_0, \tau_s \right) = \left( 38.6, 1 \ \mathrm{mm}, 50 \ \mathrm{s}, 10 \ \mathrm{s} \right)$.
Fig.~\ref{fig2}(e) shows the real part of the output VUV field Re$\left[ \Omega_2 \left( t, L/2\right) \right] $. 
For $\theta = 0$ (red-solid-filled line), the output exhibits quantum beats for $t > 70$ s, resulting from interference between emissions from two targets at different gravitational redshifts.
Rotating the system by $\theta = \pi/2$ at $t = 74.8$ s (see green upward arrow) suppresses the delayed emission (black-dashed-dotted line). 
When both targets move to the resonant altitude where $\Delta_1 = \Delta_2 =0$, freezing the phase evolution of nuclear polarizations. This suppression leads to an advanced echo control discussed later.
A full inversion with $\theta = \pi$ induces the time reversal of the nuclear dynamics (blue-dashed line) and imposes a $\pi$ phase shift on the VUV field relative to the case of $\theta = 0$.
To reveal the GFC structure, we calculate the normalized output spectrum \cite{Lin2022} from the $n$th target.
\begin{equation}\label{eq4}
S_n\left( \omega\right) =\frac{\lvert\int_{-\infty}^\infty\Omega_n(t, L/2)e^{i\omega t} dt\rvert^{2}}{\max\lvert\int_{-\infty}^\infty \Omega_1(t , -L/2)e^{i\omega t} dt \rvert^{2}} ,
\end{equation}
and show our results in Fig.~\ref{fig2}(f).
The red-solid-filled line shows $S_2\left( \omega\right)$ for constant $\theta = 0$, where two dips at $\omega = \pm 106.6 \Gamma_0$ correspond to the absorption by two targets at $z_1$ and $z_2$.
The black-dashed-dotted line depicts the result of $\theta = \pi/2$ rotation.
Since the  input VUV bandwidth is much broader than the nuclear  resonance, a Fano-type interference between them arises and offers a phase control over $S_2\left( \omega\right)$ \cite{Heeg2017, Lin2022}.
A full inversion with $\theta = \pi$ (blue-dashed line) introduces a $\pi$ phase shift in the time-delayed signal depicted in Fig.~\ref{fig2}(e), and so flips the absorption dips into peaks in Fig.\ref{fig2}(f) \cite{Heeg2017, Lin2022}.
%
%
\begin{figure}[t]
\includegraphics[width=0.48\textwidth]{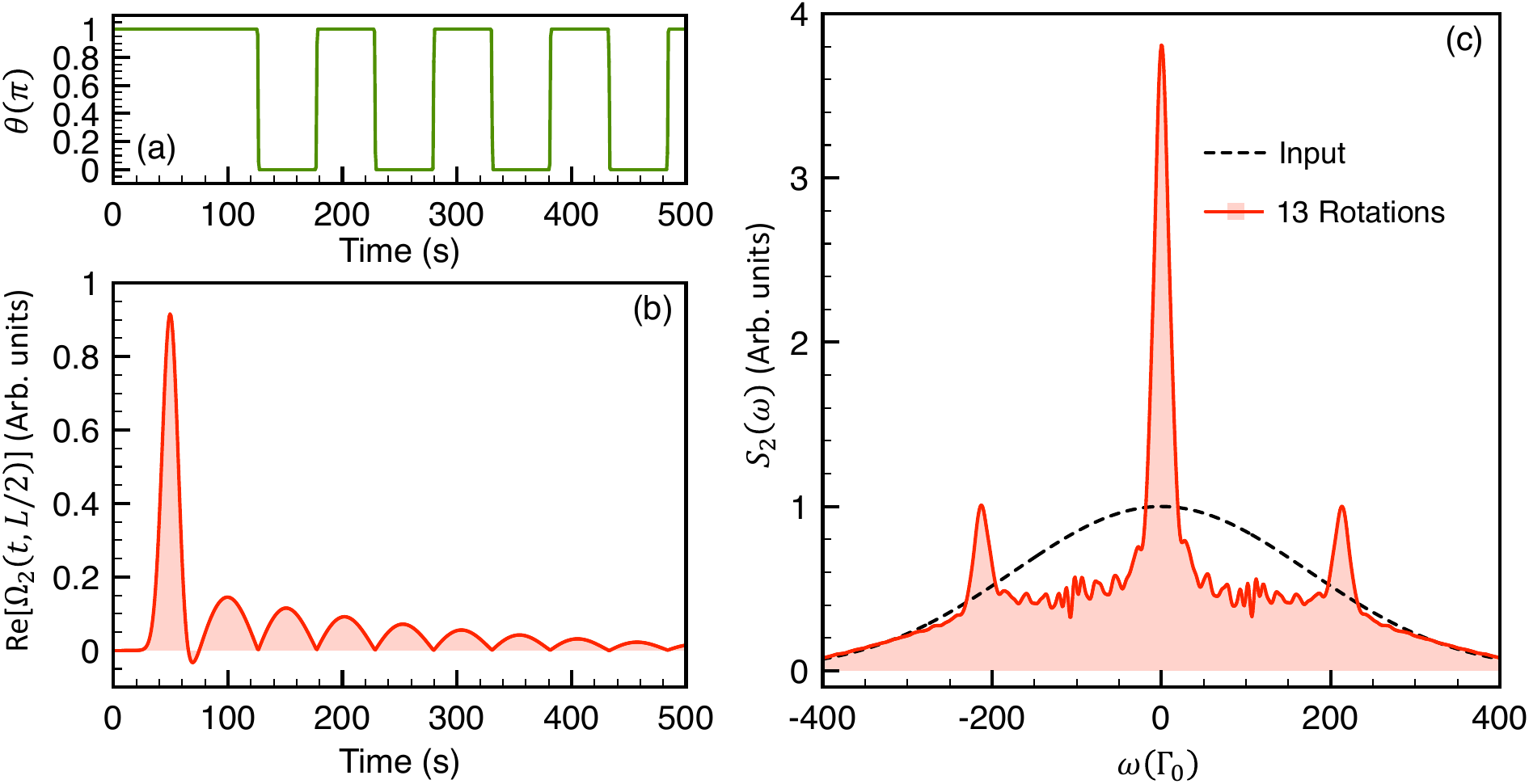}
\caption{\label{fig3}
Two $^{229}$Th targets are rotated by (a) 13 rotations $\theta\left( t\right) $ at each temporal node of (b) the output VUV field Re$\left[ \Omega_2\left( t, L/2\right) \right]$.
(c) Input spectrum (black-dashed line) and output spectrum $S_2\left( \omega\right) $ for 13 rotations (red-solid-filled line).
}
\end{figure}
%
Further manipulation of $S_2\left( \omega\right)$ is possible via multiple rotations.
We suggest using our rotational method to convert a broadband input into an enhanced narrowband VUV source, namely, transferring off-resonant photons to the resonant frequency \cite{Heeg2017, Lin2022}.
Fig.~\ref{fig3}(a) shows $\theta\left( t\right)$ which inverts the system at the first 13 temporal nodes of the output VUV field without any rotation.
As shown in Fig.\ref{fig3}(b), these multiple inversions flip nearly all time-delayed fields Re$\left[ \Omega_2\left( t, L/2\right) \right] $ to the positive value in contrast to the usual oscillatory behavior in Fig.~\ref{fig2}(e). 
This leads to spectral narrowing and enhancement at $\omega = 0$, and generates side double-frequency peaks at $\omega = \pm 213.2\Gamma_0$ in $S_2\left( \omega\right)$ of Fig.~\ref{fig3}(c) \cite{Heeg2017, Lin2022}.

\begin{figure}[b]
\includegraphics[width=0.48\textwidth]{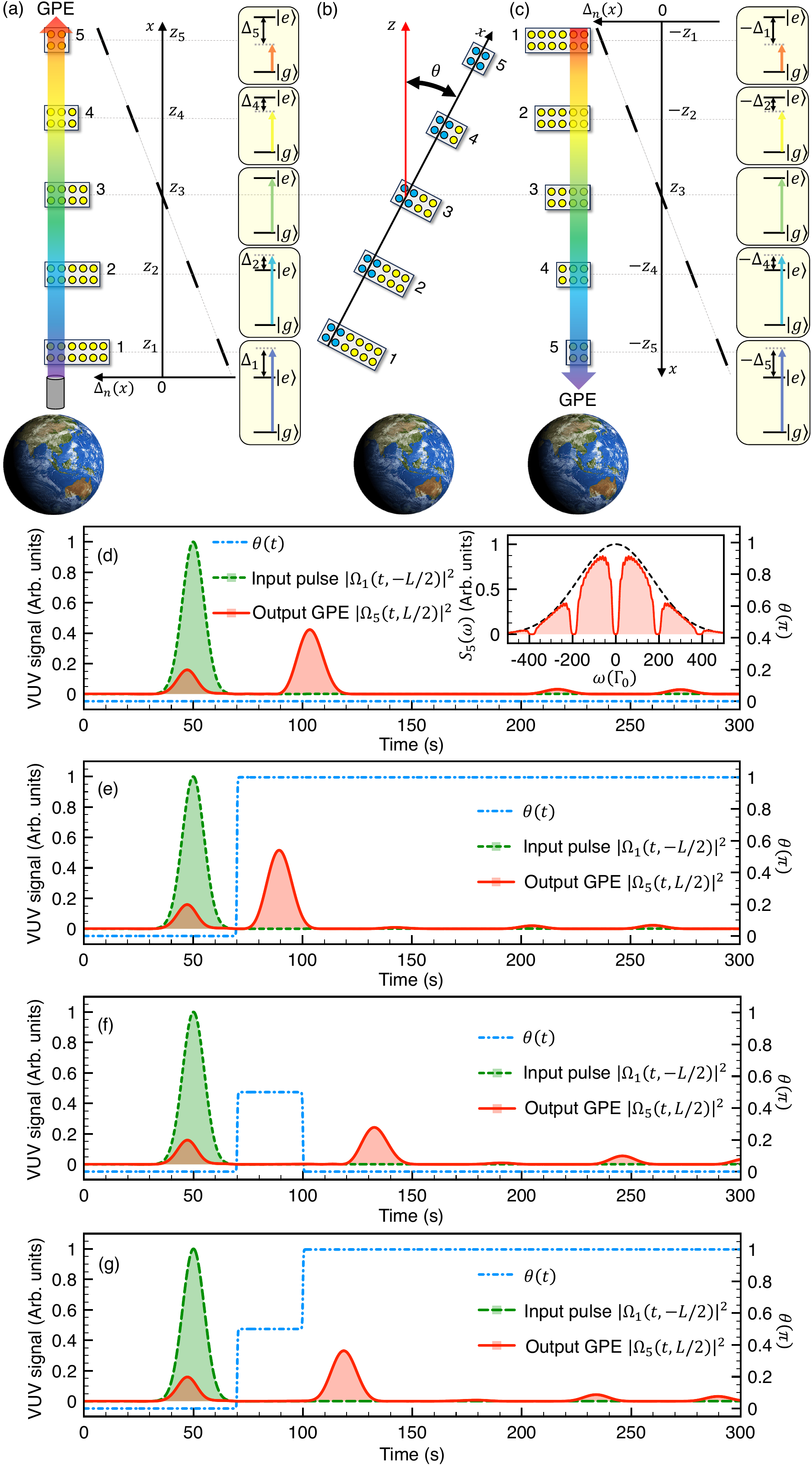}
\caption{\label{fig4}
(a) A VUV pulse (rainbow-colored upward arrow) upward illuminates five equally spaced $^{229}$Th-doped targets.
(b) The system is rotated around the center of target 3 by a polar angle $\theta$.
(c) Full inversion of the system with $\theta = \pi$.
The shape of each target indicates its orientation.
VUV output signals (blue dashed-dotted lines) under different rotation protocols:
(d) constant $\theta = 0$;
(e) monotonic rotation $\theta: 0 \rightarrow \pi$;
(f) back-and-forth rotation $\theta: 0 \rightarrow \pi/2 \rightarrow 0$;
(g) halftime rotation $\theta: 0 \rightarrow \pi/2 \rightarrow \pi$.
The inset in (d) shows the output spectrum $S_5$ without rotation, illustrating the GFC structure.
}
\end{figure}
Figures~\ref{fig2} and \ref{fig3} suggest that target rotation under gravity allows for the phase control of $\rho_{eg}$ and the coherent VUV emission. This capability offers a solution to the thickness-bandwidth limitation by forming a GFC from an array of equally spaced targets.
To resolve individual nuclear absorption lines, adjacent targets must be spaced farther apart than their thickness.
Figure~\ref{fig4}(a) illustrates a five-target system, with target shapes aiding visual identification.
An upward VUV pulse undergoes different gravitational redshifts at each target's altitude level. We tune the VUV pulse to resonance at $z_3$, namely, $\Delta_3 = 0$.
Unlike the single-target GGPE scheme, a GFC inherently produces GPE via periodic spectral beating \cite{Zhang2019}, requiring no rotation.
However, rotating the system around $z_3$, illustrated by Fig.\ref{fig4}(b), modifies both the gradient and  $\Delta_n$ and offers additional control flexibility.
Figure\ref{fig4}(c) shows the fully inverted case of $\theta = \pi$.

We study the five-target GFC  by numerically solving Eq.~\eqref{eq1} and Eq.~\eqref{eq2} with parameters $\left( \xi, L, t_0, \tau_s \right) = \left( 241, 0.1 \ \mathrm{mm}, 50 \ \mathrm{s}, 10 \ \mathrm{s} \right) $ and the initial altitude of each target center $\left( z_1, z_2, z_3, z_4, z_5 \right) = \left( -16 \ \mathrm{cm}, -8 \ \mathrm{cm}, 0 , 8 \ \mathrm{cm}, 16 \ \mathrm{cm} \right) $. The corresponding initial gravitational detunings are $\left( \Delta_1, \Delta_2, \Delta_3, \Delta_4, \Delta_5 \right) = \left( 387.6 \ \Gamma_0, 193.8 \ \Gamma_0, 0 , -193.8 \ \Gamma_0, -387.6 \ \Gamma_0 \right) $.
Fig.~\ref{fig4}(d) depicts the result without any rotation, i.e., constant $\theta = 0$ (blue-dashed-dotted line). The delay time of the first echo (red-solid-filled line) is $\tau_1 = 53.8$ s, closely matching the theoretical value $2 \pi/\left( 193.8\Gamma_0\right) = 56.4$ s. 
We evaluate the $m$th GPE signal's quality via its efficiency $R_m$ and fidelity $F_m$ \cite{Liao2014b, Zhang2019}:
\begin{eqnarray}
R_m  &=& \frac{\int_{a_m}^{b_m} \vert \Omega_5\left( t,L/2\right) \vert^2 dt}{\int_{-\infty}^\infty \vert \Omega_1\left( t,-L/2\right) \vert^2 dt},\\
F_m  &=& \frac{\vert \int_{a_m}^{b_m}  \Omega_1^\ast\left( t-\tau_m,-L/2\right) \Omega_5\left( t,L/2\right)  dt \vert^2}{\int_{-\infty}^\infty \vert \Omega_1\left( t,-L/2\right) \vert^2 dt \int_{a_m}^{b_m} \vert \Omega_5\left( t,L/2\right) \vert^2 dt}.
\end{eqnarray}
The parameters  $\left( a_m, b_m, \tau_m\right)$ for the $m$th GPE occurring in $a_m \leq t \leq b_m$ with a delay time $\tau_m$ will be specified below.
$R_m$ represents the energy conversion efficiency from the input (green-dashed-filled line) to the output (red-solid-filled line), and $F_m$ quantifies the similarity between the output echo $\Omega_5\left( t,L/2\right)$ and the input $\Omega_1\left( t,-L/2\right)$.
For the first echo, we find $R_1  = 46.44\%$ and $F_1=99.4\%$ using parameters $\left( a_1, b_1, \tau_1\right) = \left( 80  \ \mathrm{s}, 130  \ \mathrm{s}, 53.8  \ \mathrm{s}\right)$. In the  inset of Fig.~\ref{fig4}(d), the red-solid-filled line shows the five-comb structure in the output spectrum $S_5$, and black-dashed line depicts the input VUV spectrum.
Fig.~\ref{fig4}(e) displays the result for an inverted system. Time reversal rephases nuclear polarizations, advancing the first GPE signal. 
One can still observe the typical echo signature that
the input at $t = 50$ s and the GPE at $t = 90$ s appear as mirror images about the reversal time $t = 70$ s. With reduced decoherence duration for only 40 s, a stronger echo is achieved and results in $R_1 = 55.7\%$ and $F_1 = 98.8\%$ with parameters $\left( a_1, b_1, \tau_1 \right) = \left( 70 \ \mathrm{s}, 120 \ \mathrm{s}, 39.5 \ \mathrm{s} \right)$.
Based on Fig.\ref{fig2}(e), we can store and subsequently retrieve a GPE on demand by invoking  a $\theta = \pi/2$ rotation to freeze the nuclear phase evolution.
As illustrated by the blue-dashed-dotted $\theta\left( t\right) $ curve in Fig.~\ref{fig4}(f), we apply a back-and-forth $\theta = \pi/2$ rotation at $t = 70$ s and $t = 100$ s.
The nuclear dynamics is frozen for 30 s and then turned on at $t=100$ s, which delays the first GPE to $t = 133.4$ s
Due to extended decoherence, the echo weakens with $R_1 = 27\%$ and $F_1 = 93.6\%$ for parameters $\left( a_1, b_1, \tau_1 \right) = \left( 110 \ \mathrm{s}, 160 \ \mathrm{s}, 83.4 \ \mathrm{s} \right)$.
Fig.~\ref{fig4}(g) shows that a $0 \rightarrow \pi/2 \rightarrow \pi$ rotation sequence also extends the rephasing time by 30 s compared to the result in Fig.~\ref{fig4}(e), yielding $R_1 = 37\%$ and $F_1 = 97.2\%$ for the parameters $\left( a_1, b_1, \tau_1 \right) = \left( 90 \ \mathrm{s}, 150 \ \mathrm{s}, 69 \ \mathrm{s} \right)$.
Figs.~\ref{fig2}(d)–(g) demonstrate the ability to control the GPE delay time through different types of rotational operations.

The GPE system provides a platform to explore the interplay between the transverse Doppler shift (TDS) $\delta_n$  and $\Delta_n$.
The $n$th target motion with the tangential speed $v_n$ is perpendicular to the VUV propagation, and TDS $\delta_n \simeq -\frac{E_t}{2\hbar}\left( \frac{v_n}{c}\right) ^2$ will happen \cite{Jackson2021}.
We evaluate the influence of TDS by the ratio
$
r_{s} = \vert\frac{\delta_n}{\Delta_n}\vert  = \frac{v_n^2 R_E^2}{2 G M_E z_n}
$ and use $r_s = 1$ to  calculate the critical speed 
$ v_c = \frac{\sqrt{2 G M_E z_n}}{R_E}$.
Having $z_5 = 16$ cm in Fig.~\ref{fig4},  TDS is negligible when $r_{s} \ll 1$ and target's  speed is smaller than $ v_c = 1.77$ m/s.
Interestingly, $r_s = 1$ marks a region where the nuclear quantum phase induced by Earth's gravity and by
the target motion are equally significant. 
Thus, one can control the echo signal by choosing, e.g., $\left( z_5 , v_5\right) = \left( -16  \ \mathrm{cm}, 1.77 \ \mathrm{m/s} \right) $ to cancel the frequency shift $\Delta_5  + \delta_5 =0$, or $\left( z_5 , v_5\right) = \left( 16  \ \mathrm{cm}, 1.77 \ \mathrm{m/s} \right) $ to double it.

In conclusion, we explored a system of equally spaced $^{229}$Th-doped targets, where Earth’s gravitational redshift forms a GFC that generates a GPE. Rotations by $\theta = \pi$ and $\pi/2$ enable time reversal and phase freezing, allowing precise control of GPE storage and retrieval. Our approach supports applications in relativistic geodesy \cite{Flury2016} with millimeter-scale resolution via beating period or echo delay measurements, and enables enhanced narrowband VUV generation from broadband input \cite{Heeg2017, Lin2022}. This work establishes a foundation for gravitational quantum optics on Earth.

W.-T.L. is supported by the National Science and Technology Council of Taiwan (Grant No. 113-2628-M-008-006-MY3).
S.A. is supported by National  Science Foundation of China (Grant No. 11975155).
W.-T.L. conceived the idea, performed the calculation, and wrote the first draft of the manuscript.
W.-T.L. and S.A. discussed the results and improved the manuscript.

\bibliography{20250605_NFSBS2A}

\end{document}